\documentclass[longbibliography,groupedaddress,showpacs,showkeys,amssymb,eqsecnum,aps,nofootinbib,superscriptaddress]{revtex4}
\usepackage[]{graphicx}
\usepackage[]{graphics}
\usepackage{amsmath}
\usepackage{epsf}                                                                                           
\usepackage{color}                   
\usepackage{verbatim}                                                                         
\usepackage{hyperref}

\newcommand{\be}{\begin{equation}}

\newcommand{\ee}{\end{equation}}

\begin{document}                                                                                              

%%% \author{D. G. C. McKeon}
%%% \email{dgmckeo2@uwo.ca}
%%% \affiliation{
%%% Department of Applied Mathematics, The University of Western Ontario, London, Ontario N6A 5B7, Canada}
%%% \affiliation{Department of Mathematics and Computer Science, Algoma University, Sault Ste.~Marie, Ontario P6A 2G4, Canada}

%%% \author{F. T. Brandt}  
%%% \email{fbrandt@usp.br}
%%% \affiliation{Instituto de F\'{\i}sica, Universidade de S\~ao Paulo, S\~ao Paulo, SP 05508-090, Brazil}

%FTB ; JF ; DGCMK ; SMF ; GSSS.    

\author{F. T. Brandt}
\email{fbrandtl@usp.br}
\affiliation{Instituto de F\'{\i}sica, Universidade de S\~ao Paulo, S\~ao Paulo, SP 05508-090, Brazil}

\author{J. Frenkel}
\email{jfrenkel@if.usp.br}
\affiliation{Instituto de F\'{\i}sica, Universidade de S\~ao Paulo, S\~ao Paulo, SP 05508-090, Brazil}

\author{S. Martins-Filho}   
\email{sergiomartinsfilho@usp.br}
\affiliation{Instituto de F\'{\i}sica, Universidade de S\~ao Paulo, S\~ao Paulo, SP 05508-090, Brazil}

\author{D. G. C. McKeon}
\email{dgmckeo2@uwo.ca}
\affiliation{
Department of Applied Mathematics, The University of Western Ontario, London, Ontario N6A 5B7, Canada}
\affiliation{Department of Mathematics and Computer Science, Algoma University, 
Sault Ste.~Marie, Ontario P6A 2G4, Canada}

\author{G.  S.  S.  Sakoda}   
\email{gustavo.sakoda@usp.br}
\affiliation{Instituto de F\'{\i}sica, Universidade de S\~ao Paulo, S\~ao Paulo, SP 05508-090, Brazil}

\title{Thermal gauge theories with Lagrange multiplier fields}

\date{\today}

\date{\today}

\begin{abstract}
We study the Yang-Mills theory and quantum gravity at finite
temperature, in the presence  of Lagrange multiplier fields. These
restrict the path integrals to field configurations which obey the 
classical equations of motion. This has  the effect of doubling the
usual one--loop thermal contributions and of suppressing all radiative 
corrections at higher loop order. 
Such theories are renormalizable at all temperatures.
Some consequences of this result 
in quantum gravity are briefly examined.
\end{abstract}


\maketitle

\section{Introduction}
The standard theory of quantum gravity, which is based upon the
Einstein-Hilbert (EH) Lagrangian, is non-renormalizable \cite{tHooft:1974bx,Goroff:1986th,vandeVen:1991gw}.
It requires an infinite number of counterterms which involve higher-order 
terms in the curvature tensor, to cancel all the ultraviolet
divergences \cite{Gomis:1995jp,PhysRevD.100.026018}.
Many studies have been devoted to obtain a quantum field theory that
is renormalizable and unitary while having general relativity as a
classical limit. These entail, for example, the introduction of additional
terms and fields into the action, like higher derivatives models
\cite{Mannheim:2011ds} 
and supergravity \cite{Freedman:2012zz}, or the appeal to non-perturbative
properties of  renormalization group functions \cite{Nagy:2012ef}. These
attempts have met with varying degrees of success. Moreover, it has
been argued that quantum gravity may be an effective field theory
obtained in the low energy limit of a string theory \cite{green_schwarz_witten_2012,polchinski_1998}.

Recently,  it has been proposed 
that there may be an
alternative way of quantizing the EH action that removes such
divergences in a simpler manner,  preserves unitarity and produces the 
expected tree level effects 
\cite{McKeon:1992rq,PhysRevD.100.125014,Brandt:2018lbe,McKeon:2020pkm,Brandt:2021ycu}.
This involves the introduction of a
Lagrangian multiplier (LM) field that restricts the path integral used 
to quantize the theory to paths that satisfy the classical
Euler-Lagrange equations. One finds that this procedure yields,  at
zero temperature,  twice the usual one--loop contributions and that 
all higher order radiative corrections vanish. 

We are thus motivated to extend this analysis to quantum gravity with LM fields at finite 
temperature. However,  since such calculations are rather involved, we
will follow in this work Feynman' s approach \cite{Klauder:1972lsv} :
``The algebraic complexity of the gravitational field equations is so great that it is very difficult to investigate it...The Yang-Mills theory is also a non-linear theory which might show the same kind of problems, and yet be easier to handle algebraically. This proved to be the case, and thereafter all the work was done first with the Yang--Mills theory and then the corresponding expressions for gravitation were worked out.''    
Thus, in section 2, we study the Yang-Mills theory with a Lagrange multiplier field at finite
temperature
\cite{Braaten:1990mz,Frenkel:1989br,lebellac:book96,kapusta}.
To this end, we introduce the method of
forward scattering amplitudes \cite{Frenkel:1989br, Barton:1990fk} which greatly simplifies 
the computations. With these insights, we extend in section 3 the
analysis to thermal quantum gravity with Lagrange multiplier fields. 
We find that, up to one--loop order, the thermal effects are twice
those obtained in the usual theories,  while all  contributions beyond
one--loop order are eliminated.  In section 4,  we present a short
discussion of the results and their possible implications in quantum gravity.

\section{Thermal Yang-Mills theory with LM fields}

The usual Yang-Mills (YM)  theory is described by the Lagrangian
\be\label{e1}
{\cal L}_{YM}=-\frac 1 4 f^a_{\mu\nu} f^{a\mu\nu};\;\;\;
f_{\mu\nu} = 
\partial_\mu A_\nu^a - \partial_\nu A_\mu^a +
g f^{abc} A_\mu^b A_\nu^c ,
\ee
where $A_\mu^a$ is the gauge field. 
The Lagrange multiplier (LM) $\lambda_\mu^a$  field is introduced as   
\be\label{e2}
\lambda_\mu^a \frac{\partial {{\cal L}_{YM}}}{\partial A^a_\mu}
= -\lambda_\mu^a D^{ab}_\nu f^{b\mu\nu} ,
\ee
where $D^{ab}_\nu = \partial_\nu \delta^{ab} + g f^{acb} A^c_\nu$ is the covariant derivative.
The gauge fixing terms in the Feynman gauge,  together with the corresponding ghost terms, have the form 
\be\label{e3}
-\frac 1 2 (\partial\cdot A)^2  - (\partial\cdot A^a)
(\partial\cdot\lambda^a) 
-{\bar c}^a \partial\cdot D^{ab}(A) d^b
-{\bar d}^a \partial\cdot D^{ab}(A) c^b
-{\bar c}^a \partial\cdot D^{ab}(A+\lambda) c^b,
\ee
(see Eqs. (13) and (17) of \cite{McKeon:1992rq}) 
where $c$, $d$ and $\bar c$,  $\bar d$   
are respectively the ghost and anti-ghost fields. 
It may be verified that the complete 
Lagrangian is invariant under BRST transformations.          

The quadratic part of the Lagrangian is 
\be\label{e4}
{\cal L}^{(2)} = \frac 1 2
\begin{pmatrix}
A^{a\mu}, & \lambda^{a\mu}
\end{pmatrix}
\begin{pmatrix}
\partial^2 & \partial^2 \\
\partial^2 & 0 
\end{pmatrix}
\begin{pmatrix}
A^a_\mu \\ \lambda^a_\mu 
\end{pmatrix} -
\begin{pmatrix}
{\bar c}^{a}, & {\bar d}^{a}
\end{pmatrix}
\begin{pmatrix}
\partial^2 & \partial^2 \\
\partial^2 & 0 
\end{pmatrix}
\begin{pmatrix}
{c}^a \\ d^a 
\end{pmatrix}
\ee 
which leads to the following matrix propagator
\be\label{e5}
\Delta =
\begin{pmatrix}
\partial^2 & \partial^2 \\
\partial^2 & 0 
\end{pmatrix}^{-1}
=
\begin{pmatrix}
0 & 1/\partial^2 \\
1/\partial^2 & -1/\partial^2 
\end{pmatrix}.
\ee

We see that there is no $\langle A A\rangle$ 
propagator. 
There are, however,  mixed $\langle A \lambda\rangle = \langle \lambda A\rangle$ 
propagators as well as a propagator  
$\langle \lambda\lambda \rangle$ 
for the LM field. 
Moreover, all vertices have at most a single LM field. 
As a result, one cannot draw a Feynman diagram with more than one
loop.  Consequently,  no higher loop diagrams can contribute. 

The starting point of the thermal field theory is the new definition 
of an observable $O$ for a system in contact with a thermal bath at 
temperature $T$ 
\be\label{e6}
\langle O \rangle_T = Z^{-1} {\rm Tr}\left( e^{-H/T} \, O\right),
\ee
where $Z$ is the partition function\cite{lebellac:book96}. 
The Boltzmann factor  $\exp(-H/T)$
weights the occupation number of states that are accessible to the
system. 
In the imaginary time formalism, in momentum space, 
a thermal quantum
field theory in $3+1$ dimensions reduces to a 
$3$-dimensional Euclidean theory with an 
infinite summation over the Matsubara frequencies 
$Q^0_n = i 2\pi n T$ ($n=0,\pm 1, \pm 2,\dots$ or $n=\pm 1/2, \pm 3/2,
\pm 5/2$ for Bosons or Fermions respectively).

The thermal field theory is renormalizable, provided the 
zero-temperature theory is so. 
This is intuitively obvious as the thermal perturbative corrections come with 
a Bose-Einstein factor 
\be\label{e7}
N_B(\omega) = \frac{1}{e^{\omega/T}-1}
\ee
that cuts off any ultraviolet divergence. 

\subsection{Forward scattering amplitudes}

In the imaginary time formalism, the Bosonic Matsubara summations 
can be done 
using the relation \cite{kapusta}
\be\label{e8}
T \sum_{n=-\infty}^{\infty} f(Q^0_n) =
\int_{-i\infty}^{i\infty} \frac{dQ^0}{2\pi i} \frac{f(Q^0)+f(-Q^0)}{2}
+
\int_{-i\infty+\delta}^{i\infty+\delta} \frac{dQ^0}{2\pi i} 
\frac{f(Q^0)+f(-Q^0)}{e^{Q^0/T}-1}
\ee 
which allows us to separate the $T=0$ part (first term) 
from the finite-temperature contribution (second term).  

\begin{figure}[t]
    \includegraphics[scale=0.5]{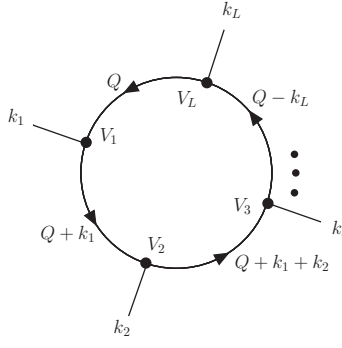}
    \caption{A generic one--loop diagram.}\label{fig1}
\end{figure}

Using Eq. \eqref{e8}, the thermal part of a generic one-loop diagram containing a number $L$
of internal lines, as shown in Fig. \ref{fig1}, is given by
\begin{eqnarray}
&&\int\,\frac{{\rm d}^3 Q}{(2\pi)^3}\int_{-i\infty+\delta}^{i\infty+\delta}
\frac{{\rm d} Q^0}{2\pi i}\, 
\frac{f(Q^0,\vec Q, k_1,k_2,\dots, k_L)+f(-Q^0,\vec Q, k_1,k_2,\dots, k_L)}{e^{Q^0/T}-1}.
\nonumber \\ &=&
\int\,\frac{{\rm d}^3 Q}{(2\pi)^3}\int_{-i\infty+\delta}^{i\infty+\delta}
\frac{{\rm d} Q^0}{2\pi i}\, \frac{1}{e^{Q^0/T}-1}
\left(f(Q^0,\vec Q, k_1,k_2,\dots, k_L)+ Q\leftrightarrow - Q
\right).
\end{eqnarray}
From Eq. \eqref{e5} we know that $f(Q^0,\vec Q, k_1,k_2,\dots, k_L)$
has the following structure
\be
f(Q^0,\vec Q, k_1,k_2,\dots, k_L) =
\frac{1}{{Q^0}^2-\vec Q^2}\frac{1}{(Q^0+k_1^0)^2-(\vec Q+\vec k_1)^2}
\cdots
\frac{t(Q, k_1,k_2\cdots k_L)}{(Q^0-k_L^0)^2-(\vec Q-\vec k_L)^2};
\;\;\; (k_1+k_2+\cdots +k_L=0).
\ee
where $t(Q, k_1,k_2\cdots k_L)$ is a tensor (or a scalar in the case
of a scalar field theory) which is determined by the interaction
vertices of the theory.  Using partial fraction decomposition like
\be
\frac{1}{(Q^0+{k}^0)^2-(\vec Q+\vec k)^2} =
\frac{1}{2\,|\vec Q + \vec {k}|}\left[
\frac{1}{Q^0+ k^0 - |\vec Q + \vec {k}|} -
\frac{1}{Q^0+ k^0 + |\vec Q + \vec {k}|}
\right]
\ee
($(k^0,\vec k)$ 
represents any of the following possibilities: $(0,\vec 0)$, $(k^0_1,\vec k_1)$, $(k^0_1+k^0_2,\vec k_1+\vec k_2)$, $\dots$, $(-k^0_l,-\vec k_l)$)
the $Q^0$ integral can be done upon
closing the integration contour on 
the right hand side of the 
%the right 
complex plane
and using Cauchy's integral formula.
Then, performing shifts in the momentum $\vec Q$ and using 
the property $\exp(x+i 2\pi m) =
\exp(x)$; $m=0,\pm 1, \pm 2, \dots$, one can show that
\be\label{barton1}
\left[
\begin{array}{l}
    \includegraphics[scale=0.5]{one_loop.eps}
\end{array}\right]_{\rm thermal}
=  -\int\,\frac{{\rm d}^3 Q}{(2\pi)^3}
\frac{1}{2|\vec Q|}\frac{1}{e^{|\vec Q|/T}-1}
%\left[ 
{\cal A}(Q,k_1,k_2,\cdots,k_L)
%+ Q\leftrightarrow  - Q
%\right]
,
\ee 
(the minus sign arises from the clockwise contour on the right hand
complex plane)
where %the amplitude is given by
\begin{eqnarray}\label{genamp}
{\cal A}(Q,k_1,k_2,\cdots,k_L)&\equiv&
\left(
\begin{array}{c}
    \includegraphics[scale=0.7]{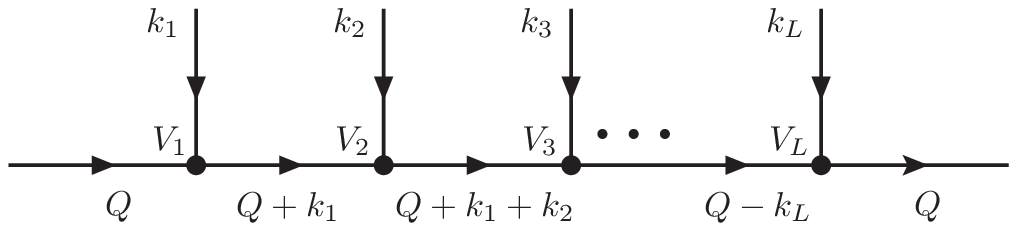}
\end{array}
+ Q\leftrightarrow  - Q
\right)_{Q^0=|\vec Q|} 
%+ Q\leftrightarrow  - Q\right\}_{Q^0=|\vec Q|} 
\nonumber \\
&&\nonumber \\
&+& 
\mbox{ cyclic permutations of } V_1,V_2,\dots,V_L .
\end{eqnarray}
The tree amplitude ${\cal A}(Q,k_1,k_2,\cdots,k_L)$ is a forward 
scattering amplitude which describes the scattering of a
on-shell particle of momentum $Q$ by external particles of momentum
$k_1$, $k_2$, $\dots$, $k_L$ \cite{Barton:1990fk}. It is
a rational function of $k^0$
which can be analytically continued to all continuous values of the
external energy.

One can now evaluate the leading  high-temperature contributions,  
which come from the hard thermal region where 
$|\vec Q|\gg |\vec k_i|,|k^0_i|$. To this end,  we expand the
denominators of ${\cal A}(Q,k_1,k_2,\cdots,k_L)$ as
\be\label{htl1}
\frac{1}{k^2 + 2 k\cdot Q} = \frac{1}{2 k\cdot Q} 
- \frac{k^2}{(2 k\cdot Q) ^2} + \dots  .
\ee
($k$ represents any of the following possibilities: $k_1$, $k_1+k_2$,
$\dots$, $-k_L$)
After combining with the contributions
from the numerator of ${\cal A}(Q,k_1,k_2,\cdots,k_L)$, the expansions
\eqref{htl1} produce terms of different degrees in $|\vec Q|$; the highest degree
yields the leading high temperature behaviour, which can be found
using the formula
\be\label{htl2}
\int_0^\infty d |\vec Q| \frac{|\vec Q|^{n-1}}{e^{|\vec Q|/T}-1} =
\Gamma(n) \zeta(n) T^n 
\ee
where $\zeta(n)$ is the Riemann zeta function.

\begin{figure}[t]
    \includegraphics[scale=0.5]{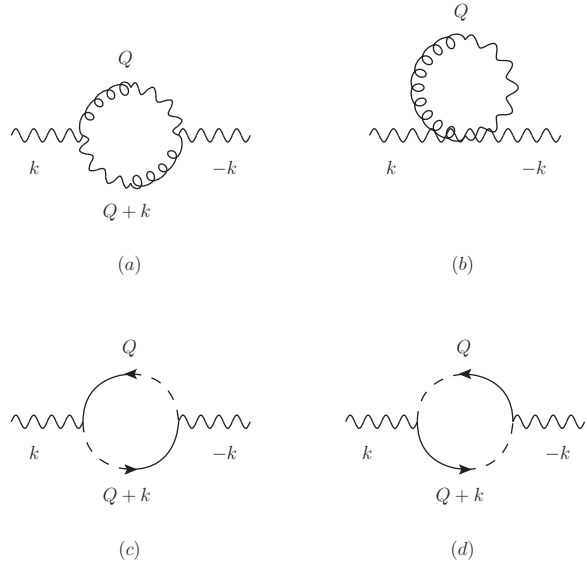}
    \caption{One-loop contribution to the thermal self-energy of the 
      gauge Boson. Springy lines represent the LM field; full lines 
      denote the $d$ ghosts and traced lines represent the $c$ ghosts.}\label{figse}
\end{figure}

\subsection{The gluon self-energy}
\begin{figure}[t!]
    \includegraphics[scale=0.55]{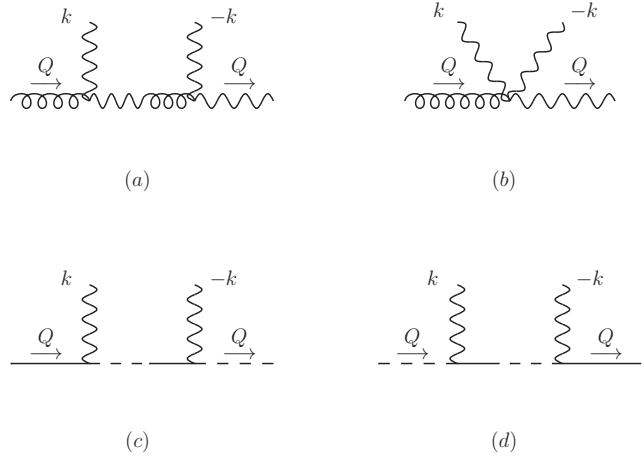}
    \caption{The forward scattering amplitudes corresponding to Fig. \ref{figse}.
Diagrams (a), (c) and (d) have to be added with
$(\mu\leftrightarrow\nu,k\leftrightarrow-k)$
as well as $Q\leftrightarrow -Q$ as indicated in the Eq. \eqref{genamp}.}
\label{figseamp}
\end{figure}

Let us consider the contributions from the gluon self-energy diagrams 
shown in Fig. \ref{figse}. These are twice those found in the usual YM theory, 
since in the later case there is a combinatorial factor of $1/2$
associated with purely gluonic graphs and only one type of ghost.  
Using Eqs.  \eqref{genamp} and \eqref{barton1},  
we can write the thermal contribution from Fig. (\ref{figse}a) as 
\be 
{\Pi^{(1)}}^{ab}_{\mu\nu}(k) = - \int \frac{d^3 Q}{(2\pi)^3}
\frac{N_B(|\vec Q|)}{2|\vec Q|} 
%\left[
{{\cal A}^{(1)}}^{ab}_{\mu\nu}(Q,k)
%+ Q\leftrightarrow -Q 
%\right]
,
\ee
where the forward scattering amplitude ${{\cal A}^{(1)}}^{ab}_{\mu\nu}(Q,k)$ is
shown in Fig. (\ref{figseamp}a) (added with the permutation $(\mu\leftrightarrow\nu,k\leftrightarrow-k)$ and $Q \leftrightarrow -Q$).

Next, let us consider the thermal contribution from the tadpole graph
shown in Fig. (\ref{figse}b). Proceeding as before, we can express it
in terms  of the forward scattering amplitude 
as indicated in Fig. (\ref{figseamp} b).          
Finally, we must consider the thermal contributions  associated with
the ghost loops shown in Figs. (\ref{figse} c) and (\ref{figse} d).           
These may also be expressed in terms of the forward scattering
amplitudes   shown in Figs. (\ref{figseamp} c) and (\ref{figseamp} d)
together with the graphs obtained by making  
$(\mu\leftrightarrow\nu,k\leftrightarrow-k, Q \leftrightarrow -Q)$.
     
Denoting by ${\cal A}^{ab}_{\mu\nu}(Q,k)$ 
the total forward scattering amplitude,  
we can express the complete thermal contribution 
%in terms of a
%momentum integral of ${\cal A}_{\mu\nu}(Q,k)$  
as
\be 
\Pi^{ab}_{\mu\nu}(k) = - \int \frac{d^3 Q}{(2\pi)^3}
\frac{N_B(|\vec Q|)}{2|\vec Q|} 
%\left[
{\cal A}^{ab}_{\mu\nu}(Q,k)
%+ Q\leftrightarrow -Q 
%\right]
.
\ee
Using the expansion of Eq. \eqref{htl1} and the formula \eqref{htl2},
we find for the leading thermal contribution from the gluon
self-energy, the expression %\input footnote2.tex 
\be\label{fsgluon}
\Pi^{ab}_{\mu\nu}(k) = \frac 1 3 g^2 C_A T^2 \delta^{ab}  
\int\frac{d\Omega}{4\pi}
\left(
 \frac{k_\mu \hat Q_\nu+k_\nu \hat Q_\mu}{k\cdot\hat Q}
- \frac{k^2}{(k\cdot \hat Q)^2} \hat Q_\mu \hat Q_\nu 
-\eta_{\mu\nu} 
\right) ,
\ee
where ${\hat Q} = (1,{\vec Q}/{|\vec Q|})$ and $C_A$ is the group factor. 
Eq. \eqref{fsgluon} satisfies the transversality condition $k^\mu \Pi^{ab}_{\mu\nu}(k) = 0$.

Finally, upon using identities for the integral over the directions $\vec Q/|\vec Q|$ 
Eq. \eqref{fsgluon} can be further simplified, yielding \cite{Brandt:1993mj}
\be 
\Pi^{ab}_{\mu\nu}(k) = \frac 2 3 C_A g^2 T^2\delta^{ab} \int\frac{d\Omega}{4\pi}
\left(k_0
\frac{{\hat Q}_\mu {\hat Q}_\nu}{k\cdot{\hat Q}} -
\eta_{\mu 0}\eta_{\nu 0}
\right) .
\ee
This result 
leads to a 
screening thermal mass
%thermal gluon mass 
squared given by 
\be 
m^2 = \frac 2 3 C_A g^2 T^2, %C_{YM}
\ee 
%where $C_A$ is the group factor, %which is twice... 
which is twice that obtained in the usual YM theory 
\cite{lebellac:book96}, as expected.

\section{Thermal Quantum Gravity with LM fields}
Next, let us consider the Einstein-Hilbert (EH) Lagrangian 
\be
{\cal L}_{EH} = \frac{1}{\kappa^2}\sqrt{-g} g^{\mu\nu} R_{\mu\nu}(\Gamma),
\ee
where $\kappa$ is related to Newton's constant $G$ 
as $\kappa=\sqrt{16\pi G}$ and the Christoffel symbol is 
\be
\Gamma^\lambda_{\mu\nu} = \frac 1 2
g^{\lambda\sigma}\left(
g_{\mu\sigma,\nu}+g_{\nu\sigma,\mu} - g_{\mu\nu,\sigma}
\right).
\ee
The tensor $R_{\mu\nu}$ is given by
\be
R_{\mu\nu}(\Gamma) = \Gamma^\rho_{\mu\rho,\nu} 
                                  -\Gamma^\rho_{\mu\nu,\rho} 
+\Gamma^\rho_{\mu\sigma} \Gamma^\sigma_{\nu\rho} 
-\Gamma^\sigma_{\mu\nu} \Gamma^\rho_{\sigma\rho}. 
\ee
For our purpose, it is convenient to expand the metric tensor        
$g^{\mu\nu}$
in terms of the deviation from the Minkowski metric       
$\eta^{\mu\nu}$
as follows 
\be
\sqrt{-g} g^{\mu\nu} = \eta^{\mu\nu}+\kappa h^{\mu\nu}.
\ee
This allows to evaluate perturbatively the thermal Green' s
functions by expanding the EH Lagrangian in powers of $\kappa$. 

The Lagrange multiplier field  
$\Lambda^{\mu\nu}$           
is introduced as 
\be
\Lambda^{\mu\nu}\frac{\partial {\cal L}_{EH}}{\partial h^{\mu\nu}}
=\frac{2}{\kappa} \Lambda^{\mu\nu} R^{l}_{\mu\nu} + {\cal O}(\kappa),
\ee
where $R^{l}_{\mu\nu}$ is the linearized Ricci tensor.

It is simpler to work in the de Donder gauge by choosing the gauge fixing Lagrangian
\be
{\cal L}_{\rm fix} = (\partial_\mu h^{\mu\nu})^2 - %\frac 1 2
\Lambda^{\mu\nu}\left(
\partial_\mu \partial_\sigma h^\sigma_{\;\nu}+
\partial_\nu \partial_\sigma h^\sigma_{\;\mu}
%+\partial^2 h_{\mu\nu}+\frac 1 2
%\left(
%\partial_\mu \partial_\nu -\eta_{\mu\nu}\partial^2
%\right) h^\sigma_{\;\sigma}
\right),
\ee
which leads to a contribution of the gravitational ghosts given by 
\begin{eqnarray}
{\cal L}_{\rm ghost} &=& 
{\bar c}_\nu\partial^2 \eta^{\mu\nu} d_\mu 
+ \kappa {\bar c}_\nu\partial_\mu\left[
h^{\mu\rho}\partial_\rho d^\nu+h^{\nu\rho}\partial_\rho d^\mu 
-\partial_\rho(h^{\mu\nu} d^\rho) 
\right]+\{\bar c\rightarrow\bar d,\;\; d\rightarrow c \}
\nonumber \\ &+&
{\bar c}_\nu\partial^2 \eta^{\mu\nu} c_\mu 
+ \kappa {\bar c}_\nu\partial_\mu\left[
(h^{\mu\rho}+\Lambda^{\mu\rho})\partial_\rho c^\nu+
(h^{\nu\rho}+\Lambda^{\nu\rho})\partial_\rho c^\mu-
\partial_\rho((h^{\mu\nu} + \Lambda^{\mu\nu}) c^\rho) 
\right].
%
%
%{\bar c}_\mu\left\{
%\partial^2\eta^{\mu\nu} + \kappa\left[
%(\partial_\rho h^{\rho\sigma})\partial_\sigma \eta^{\mu\nu}-
%(\partial_\rho h^{\rho\mu})\partial^\nu
%\right.\right.\nonumber \\ 
%&+& \left. \left.
%h^{\rho\sigma}\partial_\rho \partial_\sigma\eta^{\mu\nu}-
%(\partial_\rho\partial^\nu h^{\rho\mu})
%\right]\right\} d_\nu + \left\{\bar c \rightarrow \bar d;
%    d\rightarrow c\right\}.
\end{eqnarray}
Using a similar procedure to that employed in the YM theory,  it turns out that there is no 
$\langle hh\rangle$ propagator. 
There are, however,  mixed $\langle h\Lambda\rangle=\langle \Lambda h\rangle$ 
propagators as well as a propagator $\langle \Lambda\Lambda\rangle$ 
for the LM fields, which have the form 
\be
D{}^h_{\alpha\beta,}{}^{\Lambda}_{\mu\nu}(Q)
=-D{}^\Lambda_{\alpha\beta,}{}^{\Lambda}_{\mu\nu}(Q)=
\frac 1 2\frac{1}{Q^2} %+i\epsilon}
\left(
\eta_{\alpha\mu}\eta_{\beta\nu} + \eta_{\alpha\nu}\eta_{\beta\mu}-
\eta_{\mu\nu}\eta_{\alpha\beta}
\right).
\ee
As in the YM theory,  all the radiative corrections in quantum gravity
with LM fields vanish beyond the one--loop order.  
The Feynman diagrams which contribute to the one--loop graviton
self-energy function are shown in Fig. \ref{figse}.
The associated forward scattering amplitudes of a thermal graviton
with on-shell momenta                    are indicated in
Fig. \ref{figseamp}. As we have seen, 
such amplitudes must be multiplied by the corresponding 
Bose-Einstein factor and integrated over the 3-momentum. 
In this way, we can express the thermal self-energy graviton function as 
%yyyyyyy
\be
\Pi_{\alpha\beta,\mu\nu}(k)=
\frac{1}{4\pi^2}
\int_0^\infty dQ |\vec Q| N_B(|\vec Q|)
\int \frac{d\Omega}{4\pi}
%\left[ 
{\cal A}_{\alpha\beta,\mu\nu}(Q,k)
%+ Q\leftrightarrow -Q 
%\right]
.
\ee
The leading high-temperature limit of the forward scattering amplitude 
${\cal A}_{\alpha\beta,\mu\nu}(Q,k)$ is governed by the contributions 
quadratic in $Q$, for large values of $Q$.  
To obtain these, one needs to expand the Feynman denominators 
as shown in Eq. \eqref{htl1}. Using the formula \eqref{htl2}
we find that 
\be\label{gravseT}
{\Pi}_{\alpha\beta,\mu\nu}(k) = 
\frac{\pi^2 T^4}{60} \int\frac{d\Omega}{4\pi}
{\cal A}_{\alpha\beta,\mu\nu}(\hat Q,k),
\ee
where the forward amplitude
${\cal A}_{\alpha\beta,\mu\nu}(\hat Q,k)$
is given by the expression (${\hat Q} = (1,{\vec Q}/{|\vec Q|})$) 
\begin{eqnarray}
{\cal A}_{\alpha\beta,\mu\nu}(\hat Q,k) &=& 2\kappa^2 
\left[
\frac{k_\nu {\hat Q}_\alpha {\hat Q}_\beta {\hat Q}_\mu}{k\cdot{\hat Q}}+
\frac{k_\mu {\hat Q}_\alpha {\hat Q}_\beta {\hat Q}_\nu}{k\cdot{\hat Q}}+
\frac{k_\alpha {\hat Q}_\beta {\hat Q}_\mu {\hat Q}_\nu}{k\cdot{\hat Q}}
\right. \nonumber \\ &+& \left.
\frac{k_\beta {\hat Q}_\alpha {\hat Q}_\mu {\hat Q}_\nu}{k\cdot{\hat Q}}-
\frac{k^2 {\hat Q}_\alpha {\hat Q}_\beta {\hat Q}_\mu {\hat Q}_\nu}{(k\cdot{\hat Q})^2}
\right]
\end{eqnarray}

The expression \eqref{gravseT} is consistent  with the Ward identity 
relating the self--energy to the one--particle graviton function 
$\Gamma_{\mu\nu}$
 \cite{Brandt:1993dk}: 
\be
(2 \eta^{\alpha\lambda} k^\beta - \eta^{\alpha\beta} k^\lambda)
\Pi_{\alpha\beta,\mu\nu}(k) = -\kappa(k_\mu \Gamma_{\nu}^{\;\lambda}+
k_\nu \Gamma_{\mu}^{\;\lambda})
\ee
which follows in consequence of the invariance under general coordinate transformations. 
 
We note that the result \eqref{gravseT} yields a screening thermal graviton mass squared given by 
\be 
m_g^2 = 2(32\pi G) \frac{\pi^2T^4}{45},
%2 \kappa^2 \frac{\pi^2T^4}{45} 
\ee
which is twice that obtained  in  thermal quantum gravity
\cite{Rebhan:1990yr}.

\begin{figure}[t]
    \includegraphics[scale=0.55]{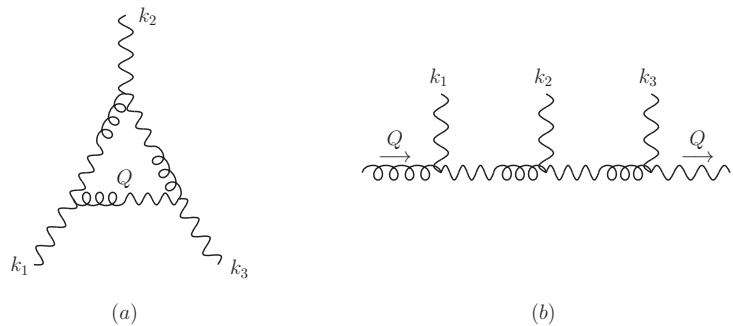}
    \caption{
Feynman diagrams contributing to the thermal three-graviton vertex 
function (a), and the associated forward scattering amplitudes (b). 
Graphs obtained by cyclic permutations of external gravitons are to 
be understood.}\label{figgrav1}
\end{figure}

One can also evaluate the leading $T^4$    
temperature corrections of the thermal three-point graviton function.  
A typical Feynman diagram is shown in 
Fig. (\ref{figgrav1}a), while the corresponding forward scattering
amplitude is indicated in Fig.  (\ref{figgrav1}b). 
This calculation is much more involved,  but the result for the total 
forward scattering amplitude 
turns out to be twice that obtained in the usual quantum gravity 
theory \cite{Brandt:1993dk}.

\section{Discussion}
We have examined the YM and quantum gravity theories with Lagrange
multiplier fields at finite temperature. We have shown that the
one--loop thermal contributions are twice those obtained in the usual
theories, with all higher--order radiative corrections being
suppressed. Such theories are renormalizable at all temperatures, 
as well as unitary. %\cite{Brandt:2021ycu}. 
% In  the present theory, the Lagrange multiplier fields are regarded as being dynamical fields. In order to ensure that this approach leads to a unitary theory with a positive  energy spectrum, the Lagrange multiplier particles must appear only in pairs \cite{Brandt:2021ycu}.This happens because in processes involving external physical gauge fields, only terms even in the Lagrange multiplier fields do occur in the intermediate states. Since the Lagrange multiplier theory reduces to Einstein's theory in the classical limit \cite{McKeon:1992rq,PhysRevD.100.125014,Brandt:2018lbe,McKeon:2020pkm,Brandt:2021ycu} it is compatible with general relativity in the low energy domain at zero temperature.
In  the present theory, the Lagrange multiplier fields are regarded as being 
dynamical fields. It can be shown that this approach leads to a unitary theory 
with a positive energy spectrum, where the Lagrange multiplier particles appear
asymptotically only in pairs \cite{Brandt:2021ycu}. Such a feature is reminiscent of that observed in the case of the strange particles, which are produced by strong interactions only pairwise. Since the Lagrange multiplier theory reduces to Einstein's theory in the classical limit at zero temperature
\cite{McKeon:1992rq,PhysRevD.100.125014,Brandt:2018lbe,McKeon:2020pkm,Brandt:2021ycu}, it is compatible with general relativity in the low energy domain.

However,  unlike the case at zero temperature,  the thermal 
zeroth order effects 
%at tree level 
also appear to be twice those obtained in the conventional
gauge theories. This may be seen, for instance, by considering the
partition function $Z$ in the Yang-Mills theory which, in the free-field
case, is given by \cite{lebellac:book96} %xxxx
\be
Z^{(0)} = \det\Delta = {\det}^2\left(\frac{1}{\partial^2}\right)=
\exp\left[2  {\rm Tr} \ln\left(\frac{1}{\partial^2}\right)\right],
\ee
where
$\Delta$ is the matrix propagator of Eq. \eqref{e5}. This leads to a
radiation pressure
\be
P = \frac{T}{V} \ln Z^{(0)}  = 2 \frac{\pi^2 T^4}{45},
\ee %zzzzzzzz
which is twice that found in the usual gauge theories. 
This fact is due to the extra degrees of freedom associated with the
LM field. This outcome depends only on the total number of degrees of
freedom, so that the same result would also be obtained in quantum gravity. 
Thus, one may inquire about some possible implications of this fact in
gravity. For example,  in a star,  it is the pressure of the
electromagnetic radiation emitted during the carbon cycle that
equilibrates the weight of the matter.  In this case,  the pressure of
the gravitational radiation is negligible.  Moreover, %We note that
in the present epoch,
the contribution of the pressure to the stress tensor for the matter
distribution on a large scale, can be %of the universe is usually 
disregarded.

The LM field interacts with gravitons, but it does not directly couple
to matter \cite{Brandt:2021ycu}.  
%The massless LM particles possess
%energy and thus can contribute to the inertia.% through their equivalent invariant mass. 
Hence, such a field exhibits a %somewhat similar
behaviour resembling that of dark matter,
which has gravitational interactions but does not otherwise couple to %but does not interact with 
ordinary matter. Thus, 
the LM field may provide an extra gravity which
%the LM particles 
could possibly affect the galactic dynamics.

%The LM field interacts with gravitons, but it does not directly couple
%to the matter particles \cite{Brandt:2021ycu}. Thus, the LM field
%behaves like a collisionless gas, so that its effect on the cosmic
%dynamics may also be neglected.
%
%Although the LM field interacts with the gravitons,  gauge invariance
%disallows a direct coupling of the LM field 
%to matter \cite{Brandt:2021ycu}. 
%Thus,  the LM field behaves like a collisionless gas which does not 
%directly interact with the matter particles. 
%Hence, its effect on the cosmic dynamics may also be neglected.  
%%%%
%The LM field interacts with gravitons but it does not directly couple 
%to matter \cite{Brandt:2021ycu}.
%Although LM particles are massless, these have energy and so might contribute 
%to the inertia through their equivalent ``invariant mass''. Thus, such a field may be
%a possible candidate for being one of the components of dark matter, which also 
%does not appear to interact with ordinary matter. Hence, LM particles could have 
%an effect on the dynamics of galaxies.
%However, at an early epoch when the universe was radiation dominated, 
%doubling the pressure of the gravitational field leads to a doubling of its
%energy density. This is a small fraction of the total energy density, so that
%the presence of the LM field may slightly modify the cosmic dynamics in 
%the early universe. This may lead to a decrease of the theoretical value of 
%the background blackbody radiation temperature in the present universe 
%by a small amount of about $1\%$ (see Appendix B).

Certainly, the only way of deciding the correctness of any
approach to quantizing gravity is to appeal to experiment.
Although such observations are not yet feasible,  there have been
several proposals for quantum gravity phenomenology.  These
include, for example,  possible observations of quantum gravity
effects in the cosmic microwave background \cite{Akhmedov2014} and 
of quantum decoherence induced by space-time fluctuations 
\cite{Oniga:2017pyq}.  
It may be worth studying further the predictions of quantum gravity
theories for such phenomena, which could be tested in the foreseeable future.

\begin{acknowledgments}
{D. G. C. M. acknowledges discussions with Roger Macleod.
%F. T. B. and  J. F.  thank CNPq (Brazil) for financial support. 
F. T. B.,  J. F.,  S. M.-F. and G. S. S. S.  thank CNPq (Brazil) for financial support.}
%This work comes as an aftermath of an original 
%project developed with the support of FAPESP (Brazil),  grant number 2018/01073-5.}
\end{acknowledgments}

%\newpage 
%\input LMThermalField_edited.bbl 

\end{document}